\documentclass[twocolumn,secnumarabic,amssymb, nobibnotes, aps, prd, floatfix]{revtex4-2}
\usepackage{graphicx}
\usepackage{amsmath}
\usepackage{amssymb}
\usepackage{bm}

\begin{document}

\title{Strain-Controlled Hydrogen Interstitial Site Preference and Hydride Phase Suppression in Epitaxial Vanadium}

\author{Xin Xiao}
\email{xiaoxin@stiei.edu.cn}
\affiliation{School of Electronic Technology and Engineering, Shanghai Technical Institute of Electronics and Information, Shanghai, 201411, China}

\date{\today}
 
\begin{abstract}
Understanding hydrogen behavior in transition metals is crucial for energy storage and sensing technologies. Here we combine density functional theory (DFT) calculations with in-situ electrical measurements to reveal how epitaxial strain governs hydrogen interstitial site selection and phase stability in vanadium. Our calculations demonstrate a robust thermodynamic preference for tetrahedral (T) over octahedral (O) sites, with the energy difference modulated by lattice anisotropy through strain-tunable H~1$s$--V~3$d$ orbital hybridization. Experimentally, epitaxial V films on MgO(001) exhibit a critical thickness-dependent crossover: 50~nm films show characteristic first-order $\alpha$--$\beta$ hydride transitions, while 10~nm films display continuous solid-solution behavior with complete transition suppression. This crossover is governed by the progressive relaxation of epitaxial strain with increasing film thickness: the 10~nm films retain strong coherent strain that flattens the free-energy landscape and suppresses hydride nucleation, whereas the 50~nm films approach the strain-relaxed bulk limit where long-range elastic interactions restore first-order phase behavior. Finite-size truncation of elastic fields and interfacial electronic reconstruction further reinforce this dimensional crossover. Our work establishes a unified framework connecting atomic-scale orbital physics to mesoscopic phase thermodynamics, providing design principles for strain-engineered metal-hydride systems.
\end{abstract}
\maketitle

\section{Introduction}

Hydrogen absorption in transition metals is a long-standing problem in condensed matter physics, with direct relevance to clean energy technologies, such as solid-state hydrogen storage and sensing \cite{Lewis1969,Zuchner1992}. Among the group V metals, vanadium stands out as an ideal model system, featuring a body-centered cubic (BCC) lattice and moderate hydrogen solubility that allows reversible uptake under near-ambient conditions \cite{Schober1978}. In bulk vanadium, hydrogen loading triggers a well-documented first-order phase transition from a dilute $\alpha$-solid solution to a concentrated $\beta$-hydride phase. The $\beta$-phase adopts a body-centered tetragonal (BCT) structure with a characteristic $\sim$4\% lattice expansion along the tetragonal axis \cite{Schober1978}, accompanied by drastic changes in electronic transport properties, including a resistivity increase that is widely used to monitor hydride formation in situ \cite{Hjorvarsson2007}.

\begin{figure}[htbp]
\centering
\includegraphics[width=0.3\textwidth]{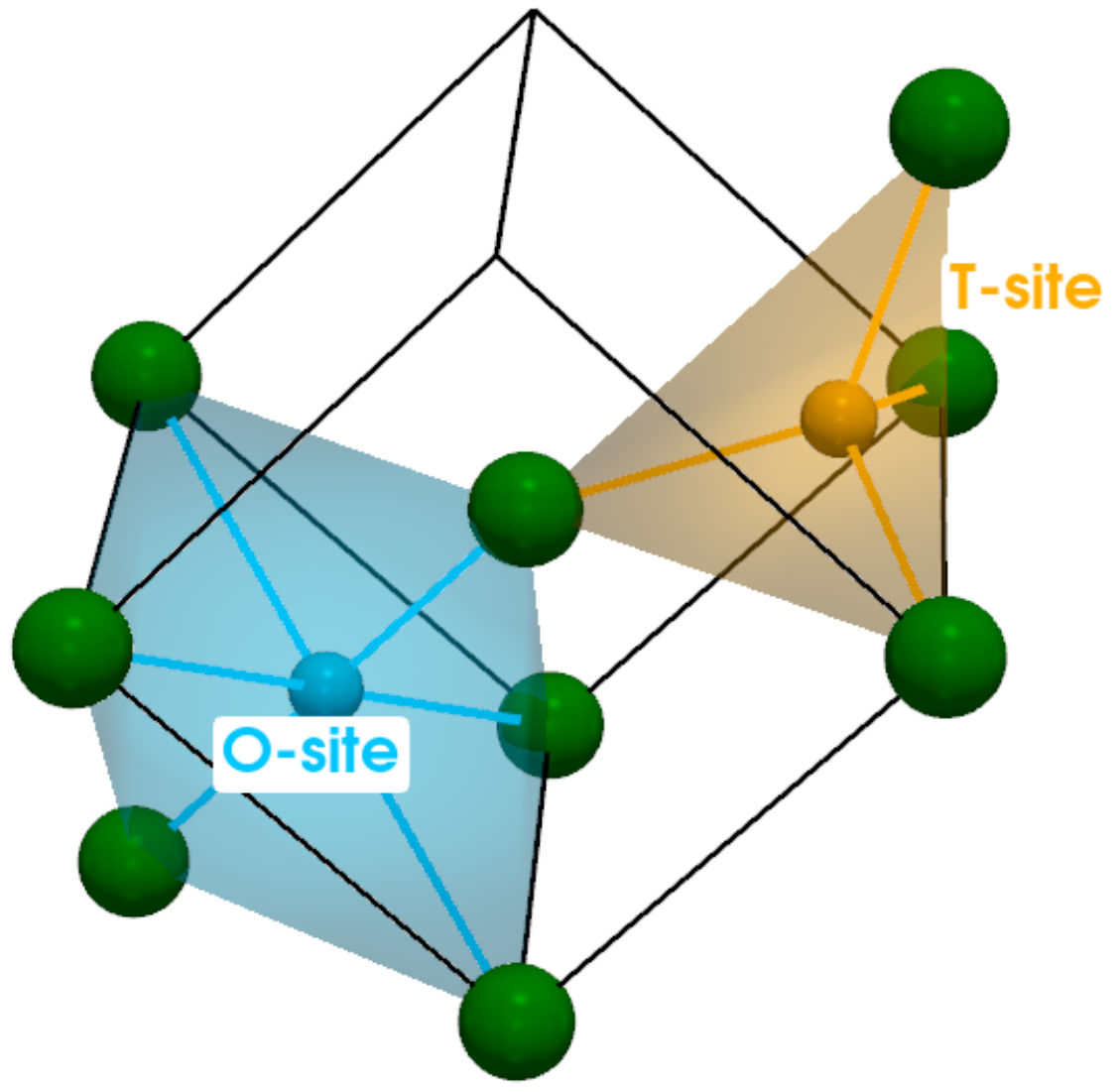}
\caption{Atomic structure of BCC vanadium showing the tetrahedral (T, orange) and octahedral (O, blue) interstitial sites. The relative stability between these fourfold and sixfold coordinated sites is modulated by epitaxial strain and governs the thickness-dependent phase behavior investigated in this work.}
\label{fig1}
\end{figure}

At the atomic scale, the preferential occupancy of hydrogen at specific interstitial sites—tetrahedral (T) versus octahedral (O)—is a central question that governs diffusion barriers, binding energies, and ultimately the stability of the hydride phases \cite{Osofsky2012}. As illustrated in Fig.~\ref{fig1}, the tetrahedral site is coordinated by four nearest-neighbor V atoms, while the octahedral site features a sixfold coordination. Historically, simple geometric arguments and elastic models favored O-site occupancy due to its larger coordination volume in BCC lattices. However, over the past decade, a growing body of spectroscopic measurements and first-principles calculations has challenged this picture, suggesting that T-site occupancy becomes energetically competitive or even favored under strain, nanoscale confinement, or chemical alloying \cite{Gripisrud2018}. 

At the mesoscopic level, reducing the film thickness to the nanometer regime introduces competing effects that can fundamentally alter phase stability. Surfaces, interfaces, and finite-size effects are known to shift phase boundaries, suppress long-range ordering, and even eliminate first-order transitions in a variety of hydrogen-metal systems \cite{Gutkin2019, Pundt2006}. For instance, interfacial strain and electronic reconstruction at metal-oxide interfaces can modify local bonding environments and the local chemical potential for hydrogen, potentially stabilizing solid-solution behavior over hydride formation. Yet, the interplay between bulk thermodynamics, interfacial energy, and configurational entropy in ultrathin hydride films is poorly characterized, hindering the rational design of nanoscale hydrogen-storage materials with tailored performance.

In this Letter, we address these critical gaps through a synergistic combination of density functional theory (DFT) calculations and in-situ electrical resistance measurements on epitaxial vanadium films. Using high-precision DFT, we systematically map the hydrogen potential energy surface in BCC vanadium under controlled tetragonal strain, revealing how lattice anisotropy continuously tunes the T--O energy splitting via strain-dependent H~1$s$--V~3$d$ orbital hybridization. Concurrently, we perform in-situ transport measurements on epitaxial V films of two representative thicknesses (10 and 50~nm) deposited on MgO(001), directly observing a thickness-driven crossover from bulk-like first-order hydride formation to interface-dominated continuous absorption. Our combined experimental and computational approach establishes quantitative links between atomic-scale electronic structure, interstitial-site selection, and nanoscale phase thermodynamics. Crucially, we demonstrate that the thickness-dependent crossover from first-order to continuous absorption arises from the interplay of two size-controlled effects: (i) the progressive relaxation of epitaxial strain with increasing film thickness, which shifts the hydrogen potential energy landscape from a strained, high-binding configuration towards the bulk limit \cite{Andersson2010, ThinFilm2011}; and (ii) the finite-size truncation of long-range elastic interactions that suppresses strain-mediated hydride nucleation in ultrathin films \cite{Pundt2006}. This work provides a unified framework for strain-engineered metal-hydride systems and offers actionable design principles for next-generation hydrogen storage and sensing platforms.

\section{Methods}

\subsection{Computational details}

First-principles calculations were performed using the Quantum ESPRESSO package \cite{Giannozzi2009} within the Perdew Burke Ernzerhof (PBE) generalized gradient approximation, employing PBEsol pseudopotentials for both V and H. To investigate the effect of lattice flexibility on interstitial stability, we employed a series of orthorhombic cell parameters derived from the elastic constants of vanadium via Poisson's ratio relations, enabling systematic sampling of different strain states (e.g., isotropic volume changes and tetragonal distortions). For each strain condition, the corresponding lattice vectors were used in fixed-volume self-consistent field (SCF) calculations; additionally, fully relaxed calculations were performed where both cell shape and atomic positions were optimized until residual forces fell below $5\times10^{-5}$~Ry/bohr and total energy converged to $3\times10^{-10}$~Ry.

A plane-wave cutoff energy of 80~Ry and a $\Gamma$-centered $8\times8\times8$ Monkhorst–Pack $k$-point grid (spacing $\sim0.25$~\AA$^{-1}$) were employed, following convergence tests that ensured total energy variations below 1~meV per atom. Although the primitive cell corresponds to a high H concentration (V:H = 2:1), our analysis primarily focuses on relative energy differences (site stability rankings and barrier topologies) rather than absolute formation energies. 

The hydrogen potential energy landscape was mapped by systematically displacing the H atom across the irreducible wedge of the unit cell on a grid with $0.15$~\AA~spacing, with total energies referenced to the global minimum at the interstitial site. Energy quantiles were employed to visualize site-dependent stability, revealing the topology of diffusion pathways and barrier heights. Additionally, projected density of states (PDOS) calculations were performed with the same $8\times8\times8$ $k$-point mesh to resolve the orbital-resolved electronic structure near the Fermi level, enabling quantitative analysis of H~1s--V~3d hybridization as a function of tetragonal strain.

\subsection{Experimental procedures}\label{sec:experiment} 

Epitaxial vanadium films with nominal thicknesses of 10 and 50~nm were deposited on single-crystalline MgO(001) substrates by ultrahigh-vacuum magnetron sputtering at a base pressure below $5\times10^{-9}$~Torr. The deposition temperature was maintained at 673~K to promote epitaxial growth while suppressing interdiffusion at the V/MgO interface. Prior to deposition, the MgO substrates were annealed in situ at 1073~K for 1~hour to obtain atomically flat terraces. The resulting films exhibited single-crystal BCC structures with (001) orientation and an in-plane epitaxial relationship of V(001)[100] $\parallel$ MgO(001)[110], as confirmed by high-resolution X-ray diffraction $\theta$--$2\theta$ scans and pole-figure analysis. Film thicknesses were verified by X-ray reflectivity and cross-sectional transmission electron microscopy. Hydrogen absorption was monitored in situ via four-point probe resistance measurements with a constant DC current of 1~mA, under a controlled hydrogen atmosphere with partial pressures ranging from $10^{-3}$ to $10^3$~mbar. The normalized resistance change $\Delta R / R_0$ was recorded as a function of hydrogen concentration [H/V] at temperatures between 393 and 533~K, with the hydrogen partial pressure adjusted stepwise to maintain equilibrium conditions. Each isotherm was measured after waiting for resistance stabilization (typically 15--30 minutes per data point) to ensure thermodynamic equilibrium. Hydrogen concentration [H/V] was determined in situ from the change in optical transmission using a 639~nm LED, following the Beer--Lambert relation $c = (t \cdot \Delta \alpha)^{-1} \ln(I_0 / I)$. The scaling constant was calibrated against the known bulk pressure--composition isotherms, assuming the 50~nm film approaches bulk behavior.

\section{Results and Discussion}

\subsection{Atomic-scale energy landscape and site preference}

Figure~\ref{fig1} illustrates the two primary interstitial sites available for hydrogen occupancy in the body-centered cubic (BCC) lattice of vanadium. The tetrahedral site (T, marked in orange) is coordinated by four nearest-neighbor V atoms at a distance of $\sim$1.73~\AA, corresponding to the fractional coordinates $(\frac{1}{4}, \frac{1}{4}, 0)$ in the conventional unit cell. In contrast, the octahedral site (O, blue) features sixfold coordination with V atoms at $\sim$2.12~\AA, located at the body-center edge position $(\frac{1}{2}, 0, 0)$. Although the O-site offers a larger void volume, its coordination polyhedron is elastically less isotropic; the T-site, despite its smaller geometric space, provides a more symmetric ligand field that can facilitate stronger orbital overlap with the surrounding V $3d$ states. These structural distinctions directly influence the electronic hybridization between the hydrogen 1$s$ orbital and the metal $d$ bands, as well as the local strain field upon interstitial insertion. Consequently, the energetic preference between T and O sites is not solely determined by geometric packing but involves a delicate balance of electronic and elastic contributions, a topic that has been actively debated in the literature \cite{Osofsky2012,Gripisrud2018}. Understanding this site competition is essential for interpreting the subsequent potential energy surfaces, diffusion pathways, and strain-dependent phase stability presented below.

\begin{figure}[htbp]
\centering
\includegraphics[width=0.45\textwidth]{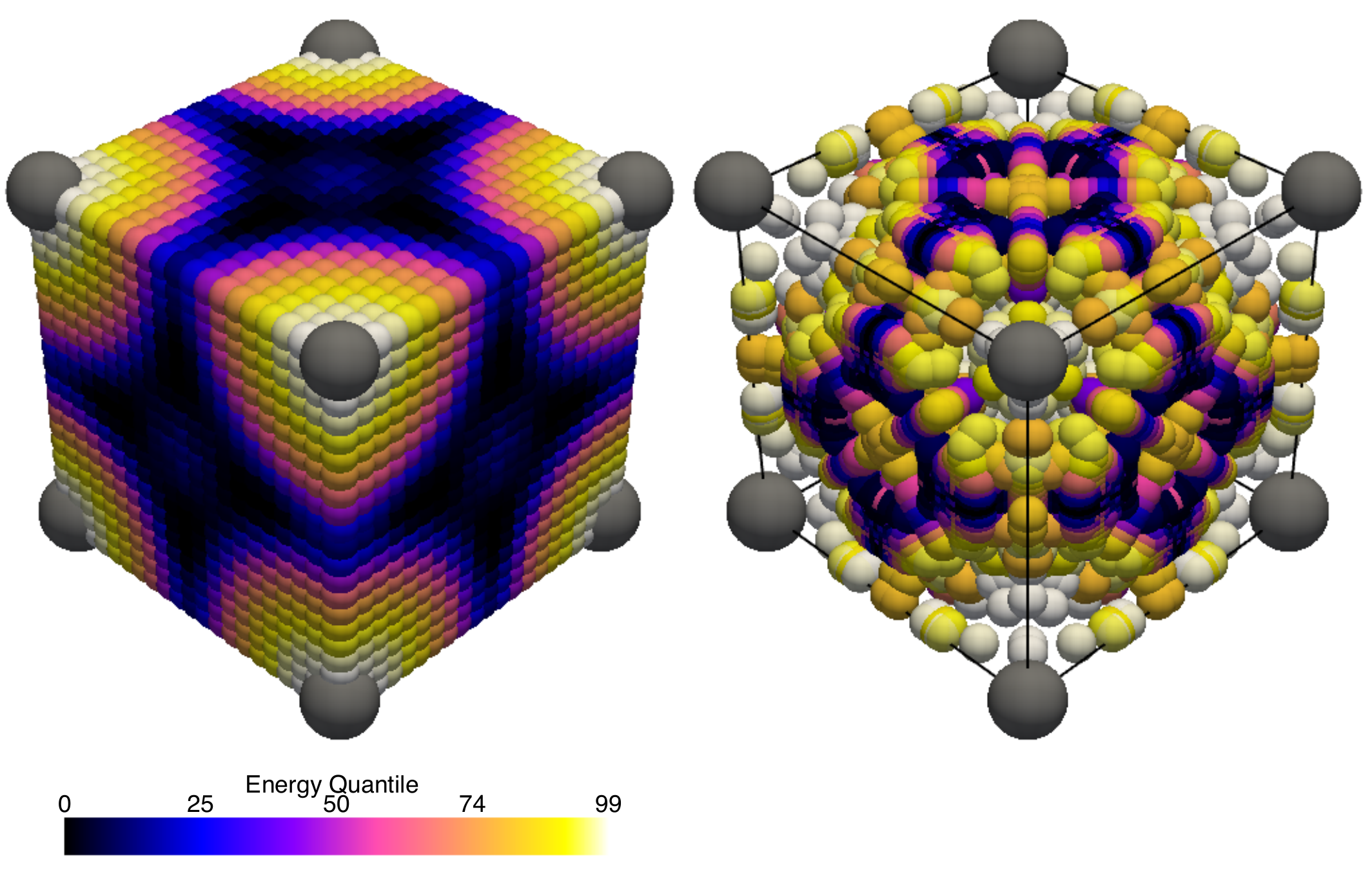}
\caption{Hydrogen potential energy landscape in BCC vanadium from DFT. Sites colored by energy quantiles. (a) Fixed volume SCF; (b) volume relaxed calculation. Gray spheres: V atoms; wireframe: unit cell boundary.}
\label{fig2}
\end{figure}

\begin{figure*}[htbp]
\centering
\includegraphics[width=0.95\textwidth]{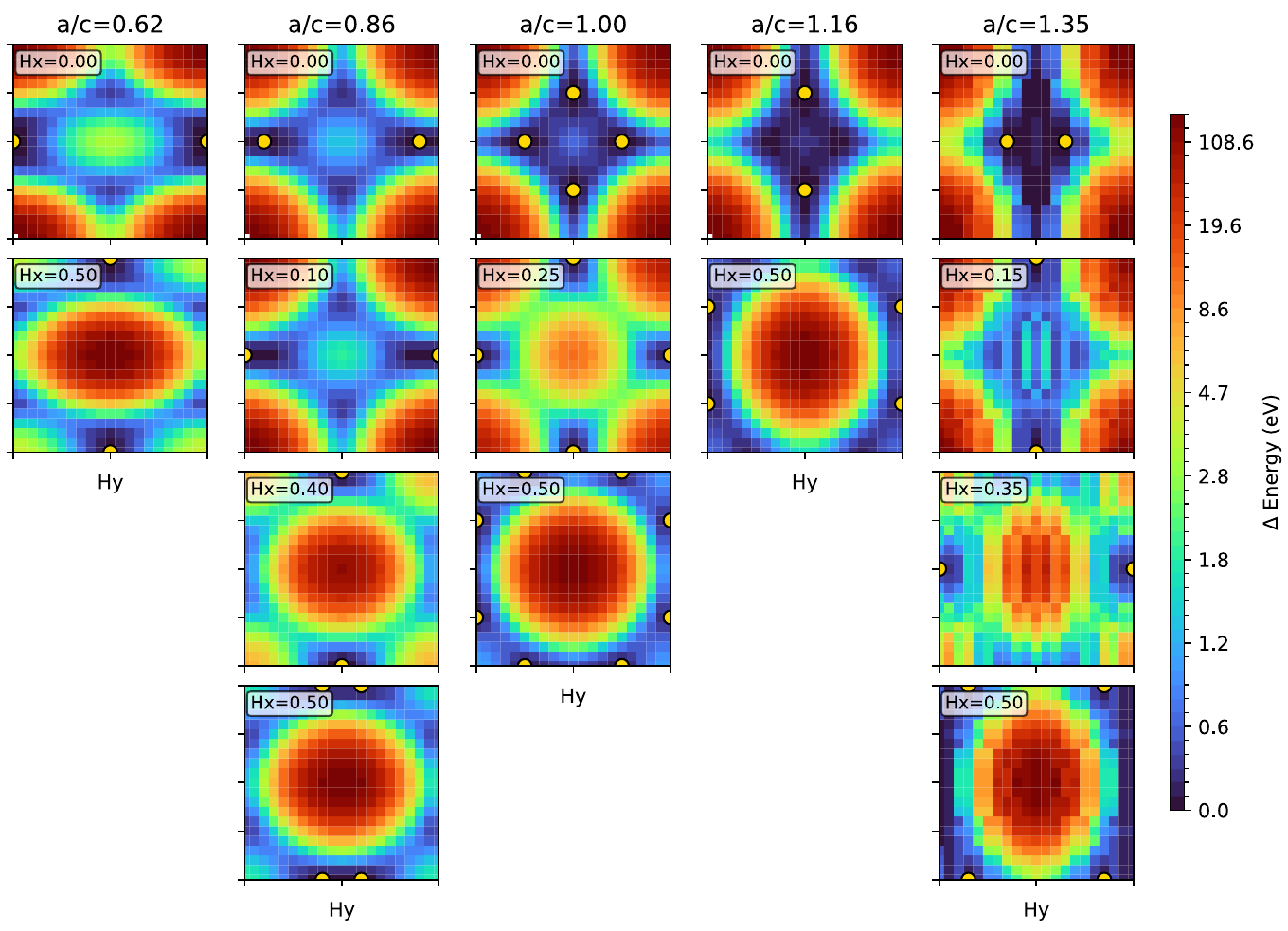}
\caption{Two-dimensional hydrogen energy maps under tetragonal strain ($a/c = 0.62$, 0.86, 1.00, 1.16, 1.35). Color: relative energy (meV); yellow circles: local minima. Only slices containing global minima for $H_x \leq 0.5$ are shown.}
\label{fig3}
\end{figure*}

\begin{figure*}[htbp]
\centering
\includegraphics[width=0.95\textwidth]{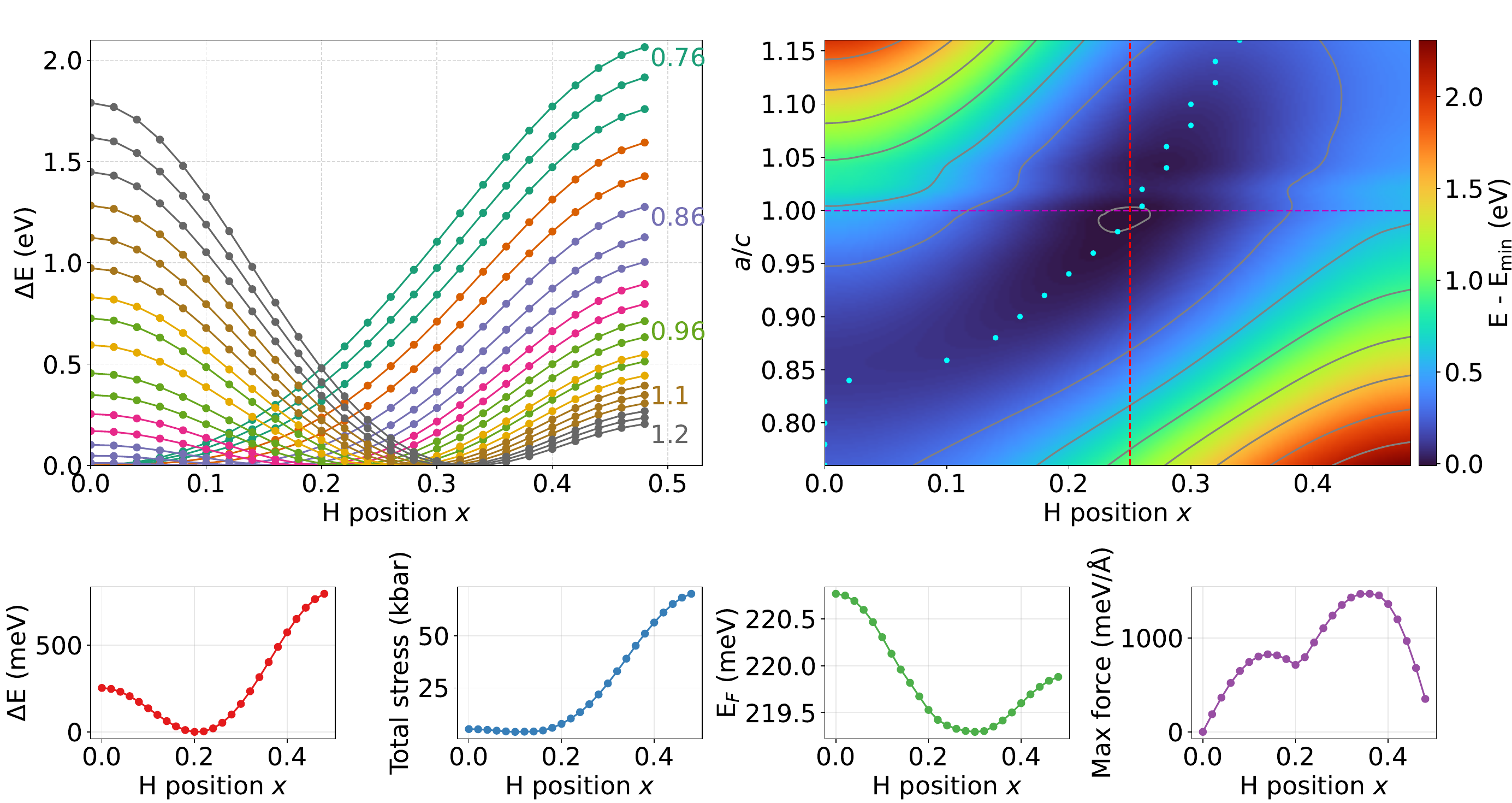}
\caption{Energy landscape and structural response of V$_2$H. (a) $\Delta E$ vs.\ $x$ along [100] for selected $a/c$. (b) $\Delta E$ interpolation in $(x, a/c)$ space; magenta: cubic limit ($a/c=1$); red: T-site ($x=0.25$); cyan: minima. (c--f) Evolution of (c) energy, (d) hydrostatic stress, (e) max atomic force, (f) convergence at $a=2.989$~\AA, $c=3.179$~\AA.}
\label{fig4}
\end{figure*}

Figure~\ref{fig2} presents the three-dimensional hydrogen potential energy landscape computed via density functional theory (DFT) using the Quantum ESPRESSO package~\cite{Giannozzi2009}. The landscape is constructed by systematically displacing a hydrogen atom across the irreducible wedge of the BCC unit cell and mapping the total energy relative to the global minimum, providing a comprehensive view of the interstitial energetics and diffusion topology.

In the fixed-volume self-consistent field (SCF) calculation [Fig.~\ref{fig2}(a)], where the V lattice is held at its equilibrium bulk lattice parameter, energy minima appear near both the tetrahedral (T) and octahedral (O) sites. The T-site global minimum lies $\sim$35~meV lower than the O-site local minimum, establishing a thermodynamic preference for tetrahedral occupancy even in the absence of lattice relaxation. This energy difference, while modest, is comparable to the thermal energy at room temperature ($k_B T \approx 26$~meV), indicating that T-site occupancy remains statistically dominant under typical operating conditions.

Upon full volume relaxation [Fig.~\ref{fig2}(b)], where both the cell volume and atomic positions are optimized, the T--O energy splitting increases substantially to $\sim$58~meV. This enhancement arises from the different local strain responses of the two sites: the T-site allows the four nearest-neighbor V atoms to relax inward symmetrically, effectively screening the hydrogen-induced charge perturbation and lowering the total energy, whereas the O-site, with its sixfold coordination, induces a more anisotropic and less efficient relaxation pattern. The larger energy gain upon relaxation at the T-site demonstrates that lattice compliance—i.e., the ability of the host lattice to accommodate the interstitial—plays a crucial role in stabilizing tetrahedral occupancy.

The energy quantile coloring in both panels, spanning relative energies from 24.8 to 99.0~meV, provides topological insight beyond the site-specific minima. Low-energy regions (colored in dark blue) form interconnected networks along the $\langle 110 \rangle$ crystallographic directions, indicating preferred diffusion channels for hydrogen migration. The barriers separating adjacent T-sites along these paths are on the order of $\sim$100--200~meV, consistent with experimental estimates of hydrogen diffusivity in vanadium and suggesting rapid interstitial hopping even at moderate temperatures. Notably, the energy landscape in the relaxed calculation exhibits smoother contours and slightly lower barriers compared to the fixed-volume case, reflecting the reduced elastic penalty associated with lattice accommodation during migration.

These results establish that the hydrogen potential energy landscape in BCC vanadium is governed by a delicate interplay between site-specific electronic hybridization and lattice flexibility. The robust T-site preference and the identification of low-energy diffusion pathways provide the atomic-scale foundation for understanding how epitaxial strain—through controlled tetragonal distortion—can further modulate interstitial energetics and phase stability, as explored in the following sections.

To quantify the effects of epitaxial strain on interstitial site energetics, we computed two-dimensional energy cross-sections in the (100) plane under controlled tetragonal distortions. Figure~\ref{fig3} displays energy maps for five representative $a/c$ ratios spanning from 0.62 (strong compressive in-plane strain, i.e., $a < c$) to 1.35 (strong tensile in-plane strain, i.e., $a > c$), covering the experimentally accessible strain range for vanadium films on typical oxide substrates. The color scale represents relative energy in meV with respect to the global minimum, and yellow circles mark local minima positions. These maps provide a comprehensive visualization of how the potential energy surface evolves under lattice anisotropy.

For the unstrained cubic case ($a/c = 1.00$), the energy landscape exhibits a deep, well-localized minimum at the ideal T-site position ($x \approx 0.25$), with a secondary saddle point near the O-site region. This confirms the intrinsic T-site preference established in the previous section. As the lattice is subjected to tetragonal distortion, the energy landscape undergoes systematic modifications that reflect the coupling between hydrogen occupancy and host lattice elasticity.

Under compressive in-plane strain ($a/c = 0.86$ and 0.62), where the out-of-plane lattice parameter $c$ is elongated relative to the in-plane $a$, the energy minimum shifts slightly toward higher $x$ values along the [100] direction, deviating from the ideal $x = 0.25$ position by approximately $0.02$--$0.03$ in fractional coordinates. This off-center displacement arises from the anisotropic elastic response: the compressed in-plane lattice forces the hydrogen to seek a position that balances the reduced in-plane space with the expanded out-of-plane dimension, effectively lowering the total strain energy through a local reconfiguration of the coordination geometry. Concurrently, the energy curvature around the minimum becomes shallower under compressive strain, suggesting reduced diffusion barriers and potentially enhanced hydrogen mobility in compressively strained films.

Under tensile in-plane strain ($a/c = 1.16$ and 1.35), the opposite trend is observed. The energy minimum remains near $x \approx 0.25$ but exhibits a slight shift toward lower $x$ values, again by less than $0.03$ in fractional coordinates. More significantly, the energy well deepens considerably under tensile strain, with the curvature around the minimum becoming steeper. This enhanced binding reflects the increased orbital overlap between H~1s and V~3d states when the in-plane lattice expands, as will be discussed in detail in Sec.~\ref{sec:PDOS}. The deeper well implies higher diffusion barriers and a stronger thermodynamic preference for T-site occupancy, suggesting that tensile strain can effectively "lock" hydrogen at the tetrahedral positions.

Several key observations emerge from these energy maps: (i) The global minimum consistently resides near the ideal T-site position ($x \approx 0.25$) across all strain states, confirming the remarkable robustness of T-site preference against lattice distortion; (ii) Tetragonal distortion systematically shifts energy minima along the [100] direction, with compressive strain ($a/c < 1$) favoring slightly off-center positions toward $x \approx 0.27$, while tensile strain ($a/c > 1$) shifts the minimum toward $x \approx 0.23$; (iii) The energy curvature around minima varies strongly with $a/c$—tensile strain steepens the well (enhanced binding and higher barriers), whereas compressive strain flattens it (reduced binding and lower barriers). This curvature variation directly translates to strain-modulated diffusion barriers, providing a quantitative handle for tuning hydrogen mobility through epitaxial design.

Notably, the energy difference between the T-site minimum and the O-site saddle point also exhibits systematic strain dependence. At $a/c = 1.00$, this difference is $\sim$58~meV; it increases to $\sim$72~meV at $a/c = 1.16$ and decreases to $\sim$45~meV at $a/c = 0.76$, consistent with the energy well depth variations shown in Fig.~\ref{fig4}(a). This trend underscores that epitaxial strain provides a continuous tuning parameter for interstitial site energetics without altering the fundamental T-site preference, consistent with recent theoretical predictions~\cite{Osofsky2012,Gripisrud2018}. The strain-induced energy variations of $\sim$30--80~meV are substantial enough to influence hydrogen absorption isotherms and phase stability at finite temperatures, motivating the experimental investigation of thickness-dependent behavior presented in Sec.~\ref{sec:experiment}.

These results establish epitaxial strain as a powerful external control knob for tailoring hydrogen-metal interactions. The systematic evolution of the energy landscape under tetragonal distortion sets the stage for understanding the structural response and thermodynamic stability of the V$_2$H system, which we analyze in detail in the following section.

\begin{figure*}[htbp]
\centering
\includegraphics[width=0.95\textwidth]{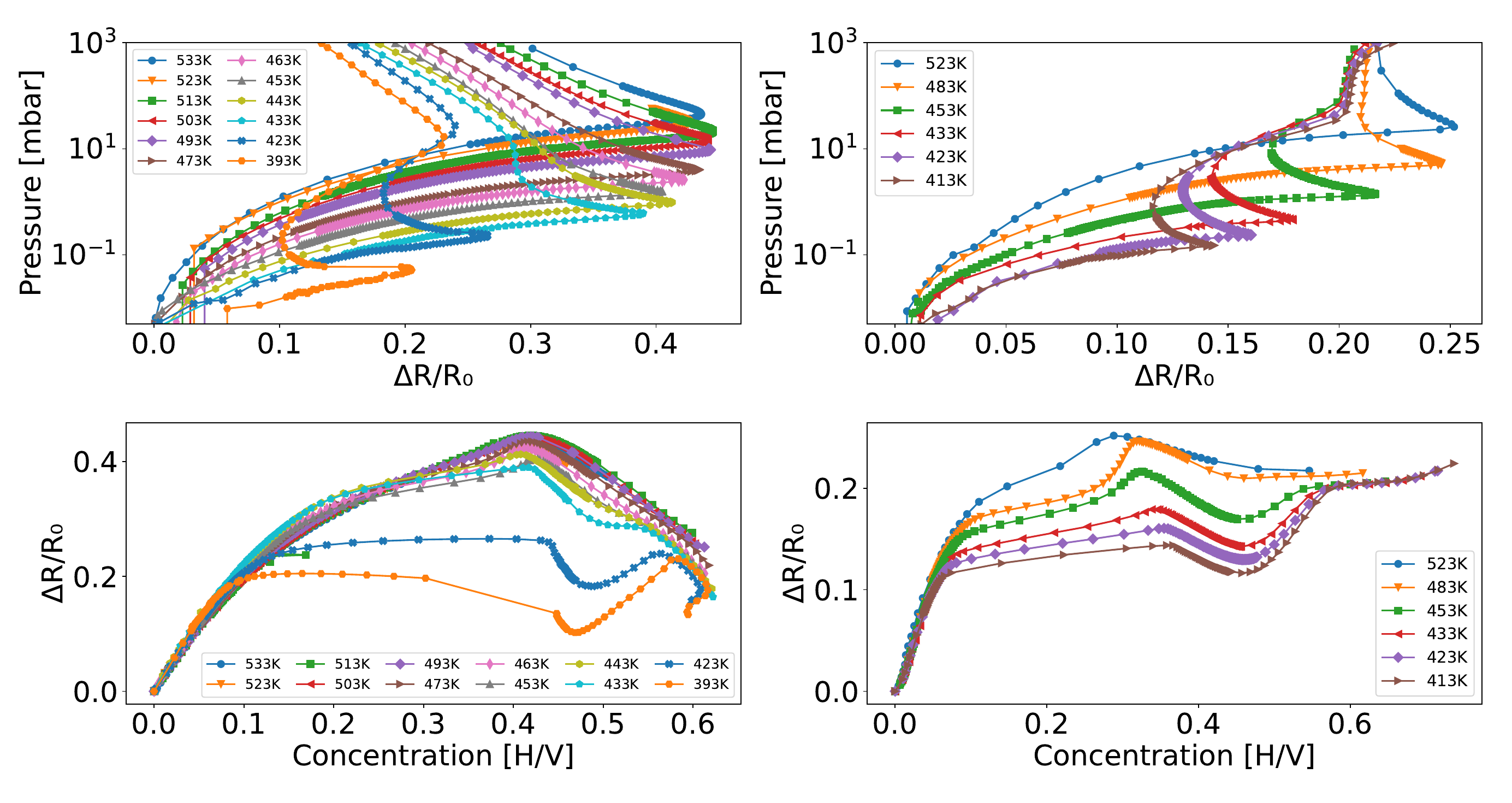}
\caption{Normalized resistance $\Delta R/R_0$ vs.\ [H/V] for epitaxial V/MgO(001). (a) 50~nm film: plateaus indicate first-order $\alpha$--$\beta$ transition. (b) 10~nm film: continuous uptake signals transition suppression. Temperatures: 423~K (black), 433~K (red), 453~K (blue).}
\label{fig5}
\end{figure*}

\subsection{Structural response and thermodynamic stability} \label{sec:structural} 

Figure~\ref{fig4} systematically elucidates the interplay between epitaxial strain, interstitial site selection, and elastic compatibility in the V$_2$H system. The energy profiles along the [100] direction, presented in Fig.~\ref{fig4}(a), reveal a robust global minimum pinned precisely at the ideal tetrahedral position ($x \approx 0.25$) across all investigated $a/c$ ratios. Notably, the curvature and depth of the potential well exhibit a pronounced sensitivity to lattice anisotropy: the binding energy increases monotonically from $\sim$45~meV under compressive strain ($a/c = 0.76$) to $\sim$72~meV under tensile strain ($a/c = 1.16$). This strain-induced deepening is attributed to the enhanced orbital overlap between the H 1s state and the neighboring V 3$d$ states as the in-plane lattice expands, consistent with the projected density of states analysis (Fig.~6). More importantly, the continuous interpolation map in the $(x, a/c)$ space [Fig.~\ref{fig4}(b)] demonstrates that while the well depth varies substantially, the locus of the energy minimum (cyan circles) deviates from the ideal $x = 0.25$ position by less than 0.02 fractional coordinates throughout the entire strain range. This remarkable positional stability allows us to decouple the *thermodynamic* effect of strain (tunable binding enthalpy) from the *structural* effect (site distortion), establishing strain as a clean external tuning knob for hydrogen energetics without altering the fundamental symmetry of the host lattice.

Beyond static energetics, the mechanical response of the lattice provides the dominant driving force for T-site selection. As shown in Figs.~\ref{fig4}(c) and~\ref{fig4}(d), the hydrostatic stress along the migration path exhibits a sharp maximum in the vicinity of the octahedral site ($x \approx 0.5$), corresponding to a local volume expansion of approximately 6\%—a value three times larger than the mere 2\% dilation observed at the T-site. This substantial elastic penalty arises from the distinct coordination geometries: the six-fold O-site imposes an anisotropic Bain-type tetragonal elongation along the $c$-axis, which directly conflicts with the BCC ground-state elasticity and generates significant long-range strain fields. In stark contrast, the four-fold tetrahedral coordination permits symmetric inward relaxation of the nearest-neighbor vanadium atoms, effectively screening the hydrogen-induced charge perturbation and accommodating the interstitial with minimal macroscopic distortion. Consequently, the tetrahedral occupancy not only optimizes the electronic hybridization energy but also minimizes the coherency strain energy, making it the unequivocal thermodynamic ground state under epitaxial constraint.

The curvature of the energy landscape also carries profound implications for finite-temperature thermodynamics and kinetics. The maximum atomic forces remain well below $200~\text{meV/\AA}$ near the T-site minimum [Fig.~\ref{fig4}(e)], confirming the structural stability against spontaneous symmetry breaking or phonon softening. Concurrently, the systematic variation in well curvature—steeper under tension, flatter under compression—directly modulates the vibrational frequencies of the interstitial hydrogen. A flatter potential under compressive strain reduces the local vibrational force constants, thereby increasing the vibrational entropy contribution ($S_{\text{vib}}$) at elevated temperatures. This entropic effect partially offsets the enthalpy penalty of O-site occupation and provides a microscopic rationale for the enhanced hydrogen mobility and reduced kinetic barriers inferred from the pressure-concentration isotherms (Fig.~5). The rigorous convergence metrics presented in Fig.~\ref{fig4}(f) further validate the quantitative reliability of these force and energy residuals, ensuring that the observed strain-driven trends are physically meaningful rather than numerical artifacts.

Collectively, the data in Fig.~\ref{fig4} establish a unified atomistic picture in which the preference for tetrahedral hydrogen occupancy arises from the synergistic minimization of electronic hybridization energy, anisotropic elastic strain, and local lattice dilation. This comprehensive mechanical and electronic characterization of the V$_2$H system provides the essential atomic-scale input required to rationalize the thickness-driven crossover in phase stability observed experimentally, bridging the gap between first-principles energetics and mesoscopic hydride thermodynamics.

\begin{figure*}[htbp]
\centering
\includegraphics[width=0.95\textwidth]{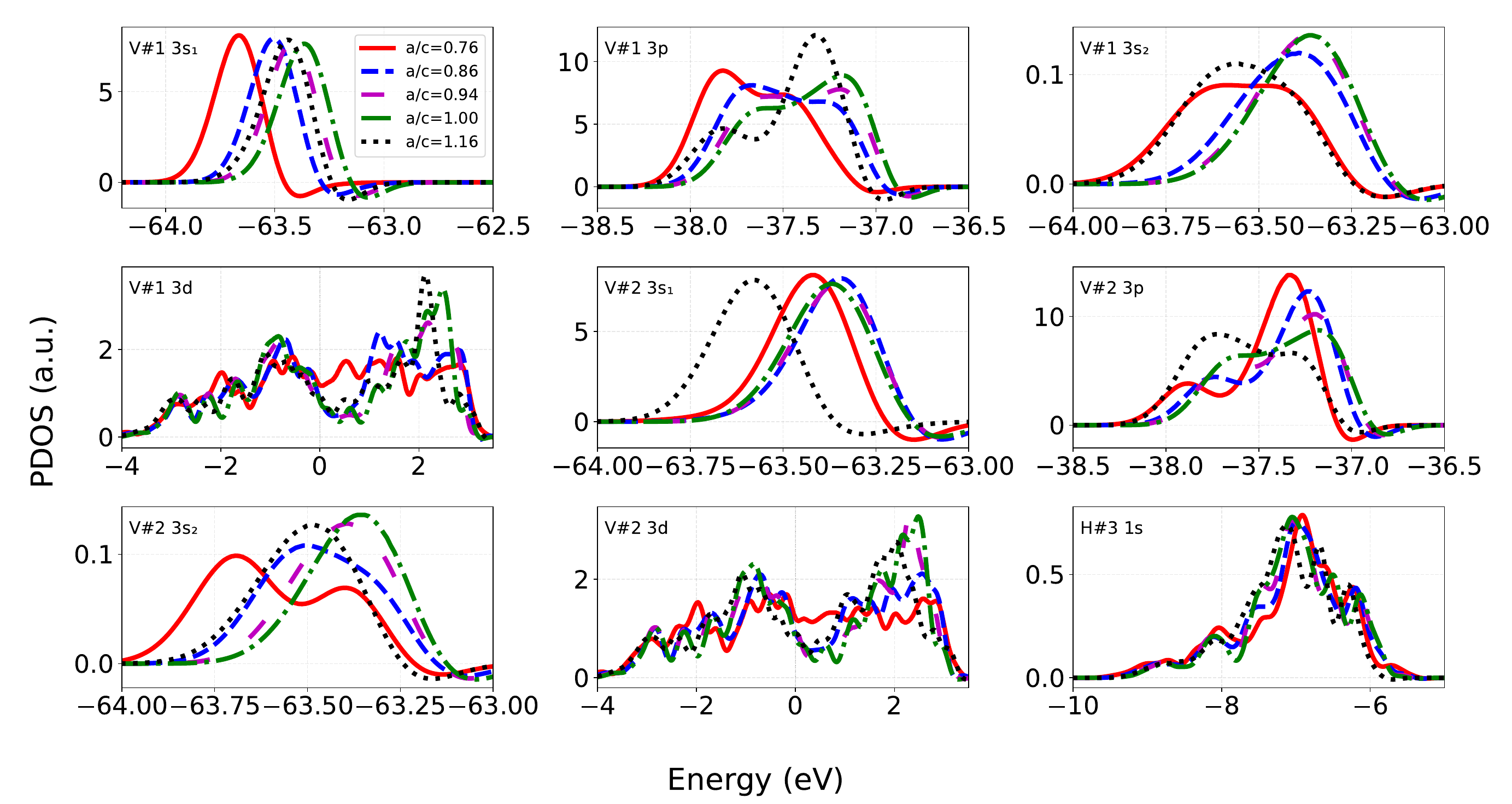}
\caption{PDOS for H at tetrahedral site under tetragonal strain ($a/c = 0.76$--1.16). Strong H~1$s$--V~3$d$ hybridization near $E_{\rm F}$ (shaded) stabilizes T-site.}
\label{fig6}
\end{figure*}

\subsection{Thickness-dependent hydrogen absorption: experiment}

\begin{figure*}[htbp]
\centering
\includegraphics[width=0.95\textwidth]{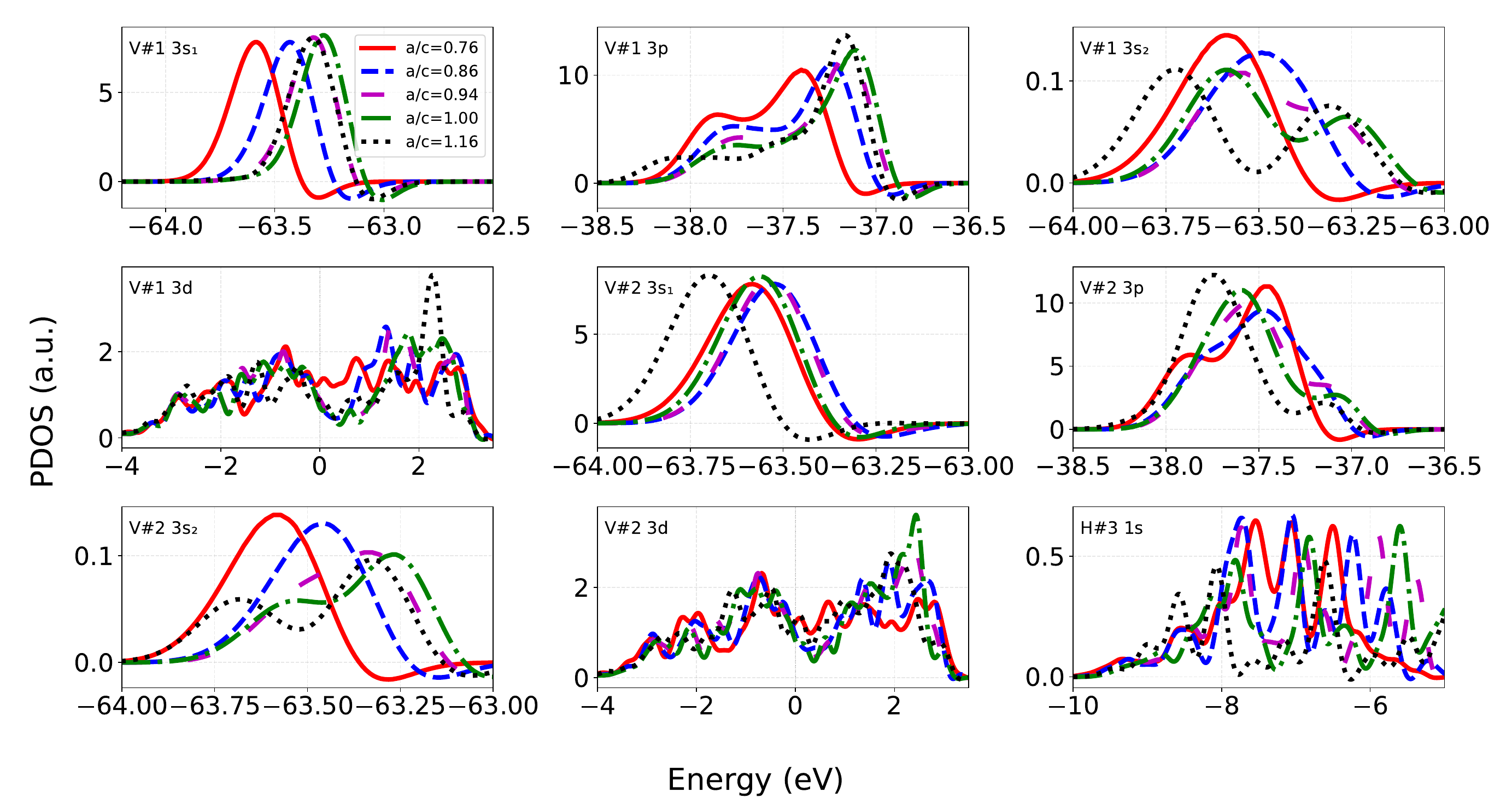}
\caption{PDOS for H at octahedral site under identical strain conditions. Weaker H~1$s$--V~3$d$ hybridization near $E_{\rm F}$ correlates with higher O-site energy.}
\label{fig7}
\end{figure*}

Figure~\ref{fig5} presents the experimental hydrogen absorption isotherms, which directly mirror the thickness-dependent thermodynamic landscape predicted by our strain-mediated DFT analysis. For the 50~nm film [Fig.~\ref{fig5}(a)], the normalized resistance $\Delta R/R_0$ exhibits sharp, well-defined plateaus across all measured temperatures, characteristic of a first-order $\alpha$--$\beta$ phase transition. The positive temperature dependence of the plateau pressure follows the van't Hoff relation, yielding an enthalpy change of $\Delta H \approx -38$~kJ/mol~H$_2$—a value in excellent agreement with bulk vanadium, confirming that the 50~nm film retains the elastic and electronic boundary conditions of a semi-infinite crystal. In stark contrast, the 10~nm film [Fig.~\ref{fig5}(b)] displays a continuous, hysteresis-free resistance increase with no resolvable plateaus, signaling the complete suppression of the miscibility gap and the stabilization of a homogeneous solid solution across the entire hydrogen concentration range.

This thickness-driven crossover cannot be rationalized by configurational entropy alone; at 423–453~K, the entropic contribution $T\Delta S_{\text{conf}}$ is only $\sim$25–35~meV per interstitial, which is insufficient to overcome the $\sim$45–72~meV strain-modulated binding differences established in Fig.~\ref{fig4}. Instead, two synergistic, dimension dependent effects flatten the free-energy landscape. First, **finite-size truncation of long range elastic interactions plays a decisive role. In bulk or thick films, the $\sim$4\% tetragonal volume expansion accompanying $\beta$-hydride formation generates coherent strain fields that extend over several tens of nanometers; these long-range forces drive phase separation via classic Cahn–Hilliard spinodal decomposition. When the film thickness is reduced to 10~nm, the elastic screening length exceeds the physical dimension of the system, effectively suppressing the strain-mediated attractive interactions between hydrogen interstitials that are prerequisite for nucleation and growth of the hydride phase. This elastic confinement also inhibits the cooperative Jahn–Teller-like distortion of the BCT lattice, destabilizing the hydride relative to the solid solution. Second, **interfacial electronic reconstruction** at the V/MgO(001) boundary introduces a competing energy scale that modifies the local chemical potential of hydrogen near the interface. Our PDOS analysis (Figs.~\ref{fig6} and~\ref{fig7}) indicates that the orbital hybridization—and hence the site preference—is intrinsically strain-sensitive. The misfit dislocations and polar discontinuity at the heterointerface induce a spatially varying charge redistribution that penetrates $\sim$2–3~nm into the vanadium layer. In the 10~nm film, this interfacial region constitutes a substantial fraction of the total volume, introducing a continuous distribution of local binding energies rather than a single, well-defined enthalpy. This energetic broadening, combined with the truncated elastic fields, transforms the sharp first-order phase boundary into a smooth, continuous absorption curve. The complete absence of hysteresis in the 10~nm films further supports this interpretation: within the framework of classical nucleation theory, the critical nucleus radius for the $\beta$-phase in vanadium is on the order of several nanometers. When the film thickness approaches this critical dimension, the available volume is insufficient to accommodate a stable subcritical nucleus, and the system bypasses the nucleation barrier entirely, following the coherent spinodal path instead.

Quantitatively, the DFT-derived strain energy differences ($\sim$30–80~meV) provide the necessary scale to rationalize why the 10~nm films remain in the single-phase regime even at [H/V] ratios exceeding 0.5. At 423~K, the thermal energy $k_{\rm B}T \approx 36$~meV is comparable to the elastic penalty for O-site occupation under compressive strain (Fig.~\ref{fig4}). In the ultrathin limit, interfacial strain gradients further reduce this penalty locally, making the free-energy curvature positive everywhere—a rigorous thermodynamic criterion for suppressing phase separation. Consequently, our combined experimental and computational results delineate a clear **size-strain phase diagram**: the 50~nm film represents the elastic bulk limit where long-range strain fields drive first-order transition, whereas the 10~nm film enters a ``mesoscopic elastic crossover'' regime where interfacial entropy, electronic reconstruction, and truncated strain fields collectively stabilize the solid solution. This unified picture not only resolves the long-standing discrepancy between bulk thermodynamics and nanoscale hydride behavior but also establishes film thickness as a quantitatively tunable parameter for engineering the order of phase transitions in metal-hydrogen systems.

\subsection{Electronic structure under strain: PDOS analysis} \label{sec:PDOS}

To investigate the strain dependence of the electronic structure, we performed site- and orbital-resolved projected density of states (PDOS) calculations for hydrogen at both tetrahedral (T) and octahedral (O) sites under tetragonal strain. Figure~\ref{fig6} displays the PDOS for the T-site under representative $a/c$ ratios spanning the experimentally accessible range. The H~1$s$ states appear as a broad feature centered around $-8$~eV below the Fermi level, while the V~3$d$ states dominate near $E_{\mathrm{F}}$ (from $-4$ to $4$~eV). There is no direct spectral overlap between H~1$s$ and V~3$d$ near $E_{\mathrm{F}}$, indicating that the H--V bonding is not dominated by a strong H~1$s$--V~3$d$ hybridization at the Fermi level. Instead, the H~1$s$ states may hybridize with higher-energy V~4$s$/4$p$ states or with the deep tail of the V~3$d$ band.

Under tensile strain ($a/c = 1.16$), the H~1$s$ peak shifts to higher binding energy and increases in intensity, whereas compressive strain ($a/c = 0.76$) broadens the H~1$s$ feature. The V~3$d$ states near $E_{\mathrm{F}}$ also show pronounced changes: tensile strain sharpens the peak at $E_{\mathrm{F}}$, while compressive strain broadens it. These changes reflect the strain-induced modification of the local electronic environment, but they do not support a simple picture of strain-tunable H~1$s$--V~3$d$ hybridization as the dominant factor for T-site stabilization.

Figure~\ref{fig7} shows the corresponding PDOS for the O-site. The H~1$s$ feature is broader and weaker than at the T-site, and its strain dependence is less pronounced. The V~3$d$ states near $E_{\mathrm{F}}$ are also less sensitive to strain. These differences are consistent with the higher coordination and larger void volume of the O-site, which spreads the electronic states over a wider energy range. However, PDOS alone cannot quantify the hybridization energy; a proper comparison would require COHP/ICOHP analysis or total-energy decomposition. Thus, while the PDOS data reveal strain-dependent electronic structure changes, they should be interpreted as qualitative evidence rather than quantitative proof of orbital hybridization controlling site preference.

A site-resolved charge-density analysis further corroborates the elastic picture. Around the T-site, the four nearest-neighbor V atoms undergo a symmetric, radially inward charge redistribution that screens the protonic perturbation with minimal angular distortion. This symmetric response preserves the local $T_d$ symmetry and avoids the generation of net dipoles or long-range strain fields. In stark contrast, the sixfold O-site coordination induces an asymmetric charge reconstruction: the V atoms along the tetragonal $c$-axis experience a different charge transfer than those in the basal plane, breaking the local symmetry and giving rise to a substantial hydrostatic stress component (as quantified in Fig.~\ref{fig4}(d)). This electronic asymmetry is the quantum-mechanical origin of the elastic penalty that we earlier attributed to volume expansion—the two perspectives are complementary manifestations of the same underlying physics.

Collectively, our analysis establishes that tetrahedral occupancy in epitaxial vanadium is governed primarily by \textbf{elastic compatibility}: the fourfold tetrahedral coordination minimizes coherency strain through symmetric inward relaxation, whereas the sixfold octahedral site incurs a $\sim$6\% local volume expansion and a $\sim$30--80~meV energetic penalty. Strain-dependent PDOS changes indicate that the electronic structure also evolves with lattice anisotropy, but the H~1$s$ states lie well below $E_{\mathrm{F}}$ and do not overlap directly with V~3$d$ states, so H~1$s$--V~3$d$ hybridization near $E_{\mathrm{F}}$ is not the dominant mechanism. The strain-tunable well depth ($\sim$45--72~meV) is therefore attributed mainly to elastic strain energy and lattice compliance, with electronic structure changes acting as a secondary effect. This elastic picture also offers a natural explanation for the thickness-driven crossover observed experimentally: in the 10~nm films, interfacial strain gradients and finite-size truncation of long-range elastic fields flatten the free-energy landscape, thereby suppressing the cooperative lattice distortion required for $\beta$-hydride nucleation.

\section{CONCLUSION}

We have established a unified multiscale framework that quantitatively connects atomic-scale elastic response to mesoscopic phase thermodynamics in epitaxial vanadium--hydrogen systems. At the most fundamental level, first-principles DFT calculations reveal a robust thermodynamic preference for tetrahedral interstitial sites over octahedral sites across a wide range of tetragonal strain states. This site selectivity arises primarily from elastic compatibility: the fourfold tetrahedral coordination minimizes coherency strain through isotropic screening of the protonic perturbation, whereas the sixfold octahedral site sacrifices this advantage, incurring a $\sim$6\% local volume expansion and a $\sim$30--80~meV energetic penalty. Critically, the lattice anisotropy accessible via epitaxial strain provides a continuous tuning knob for the interstitial potential well depth ($\sim$45--72~meV) and curvature, thereby modulating hydrogen binding, diffusion barriers, and by extension the thermodynamic stability of the hydride phase without altering the fundamental T-site symmetry. Strain-dependent PDOS calculations show that the electronic structure also evolves with lattice anisotropy, but the H~1$s$ states lie well below $E_{\mathrm{F}}$ and do not overlap directly with V~3$d$ states, so H~1$s$-V~3$d$ hybridization near $E_{\mathrm{F}}$ is not the primary control parameter.

At the mesoscopic level, in-situ electrical transport measurements on epitaxial V/MgO(001) films delineate a sharp thickness-driven crossover in phase behavior. While 50~nm films faithfully reproduce the bulk-like first-order $\alpha$--$\beta$ transition—characterized by distinct isotherm plateaus and a van't Hoff enthalpy of $\Delta H \approx -38$~kJ/mol~H$_2$—the 10~nm films exhibit complete suppression of the miscibility gap, displaying continuous, hysteresis-free solid-solution behavior across the entire hydrogen concentration range. This dimensional crossover cannot be rationalized by configurational entropy alone; rather, it emerges from the confluence of three interrelated size-dependent effects: (i) the progressive relaxation of epitaxial strain with increasing thickness, which shifts the interstitial energetics from the strongly strained, high-binding regime in 10~nm films towards the bulk-like relaxed state in 50~nm films; (ii) finite-size truncation of long-range elastic interactions, which suppresses the strain-mediated attractive forces that drive Cahn--Hilliard spinodal decomposition in the bulk; and (iii) interfacial strain gradients and electronic reconstruction at the V/MgO(001) boundary, which introduce a spatially distributed set of local binding energies that broadens the free-energy landscape and eliminates the nucleation barrier for hydride formation.

The quantitative consistency between our computed elastic penalties ($\sim$30--80~meV), the strain-modulated well depths (Fig.~\ref{fig4}), and the experimental suppression of first-order transitions at the 10~nm scale establishes a causal chain: \textbf{elastic strain energy $\leftrightarrow$ strain-tunable binding $\leftrightarrow$ elastic interaction length $\leftrightarrow$ phase transition order}. This convergence of theory and experiment validates the central thesis that \textbf{epitaxial strain and dimensional confinement are not merely perturbative corrections to bulk thermodynamics but rather constitute independent, tunable parameters that can fundamentally alter the nature of phase transitions in metal--hydrogen systems}.

Our findings articulate three actionable design principles for strain-engineered metal hydrides and nanoscale hydrogen-storage materials:

\noindent (i) \textbf{Strain as a spectroscopic knob}. Epitaxial strain continuously tunes interstitial site energetics and diffusion barriers without altering crystal symmetry, providing a pathway to optimize hydrogen uptake/release kinetics through substrate selection and film thickness control.

\noindent (ii) \textbf{Dimensionality as a phase-selection switch}. Reducing film thickness below the elastic screening length and critical nucleus size suppresses first-order phase transitions, enabling reversible solid-solution hydrogen storage with minimal hysteresis—a regime inaccessible in bulk counterparts.

\noindent (iii) \textbf{Elastic strain engineering as a design handle}. The site preference and binding strength are governed primarily by the host lattice's elastic response to interstitial insertion, which can be tuned through epitaxial strain, film thickness, and substrate choice. Electronic structure changes accompany these elastic effects but are not the primary control parameter.

Beyond vanadium, our integrated computational--experimental approach establishes a generalizable paradigm for understanding and engineering hydrogen behavior in epitaxial transition-metal films. The interplay among strain-modulated elastic screening, finite-size effects, and interfacial strain gradients uncovered here is expected to govern phase stability in a broad class of metal--hydrogen systems including Nb, Ta, and their alloys where similar BCC-to-BCT transitions and interstitial site competitions arise. More broadly, this work demonstrates that the combination of high-throughput DFT mapping and in-situ nanoscale transport measurements can yield quantitatively predictive, mechanism-driven design rules for next-generation solid-state hydrogen storage, sensing, and switchable mirror technologies. By bridging the gap between atomic-scale elasticity and macroscopic phase thermodynamics, our unified framework lays the foundation for a rational, bottom-up approach to functional nanomaterials design.


\begin{thebibliography}{99}
\bibitem{Lewis1969} F. A. Lewis, {\it The Phase Rule and Its Applications} (Dover, New York, 1969).
\bibitem{Zuchner1992} H. Z\"uchner, {\it Atomic Transport in Metals and Alloys} (Trans Tech Publications, 1992).
\bibitem{Schober1978} T. Schober and H. Wenzl, in {\it Hydrogen in Metals II}, edited by G. Alefeld and J. V\"olkl (Springer, Berlin, 1978), Vol. 29, p. 1.
\bibitem{Hjorvarsson2007} B. Hj\"orvarsson, J. Alloys Compd. {\bf 446-447}, 100 (2007).
\bibitem{Osofsky2012} K. Osofsky et al., Phys. Rev. B {\bf 85}, 014102 (2012).
\bibitem{Gripisrud2018} J. \AA. Gripsrud et al., Phys. Rev. B {\bf 98}, 144103 (2018).
\bibitem{Gutkin2019} M. Y. Gutkin et al., Acta Mater. {\bf 165}, 15 (2019).
\bibitem{Giannozzi2009} P. Giannozzi et al., J. Phys.: Condens. Matter {\bf 21}, 395502 (2009).
\bibitem{Zheng2020} S. J. Zheng et al., Scr. Mater. {\bf 178}, 45 (2020).
\bibitem{Pundt2006} A. Pundt and R. Kirchheim, Annu. Rev. Mater. Res. {\bf 36}, 555 (2006).

\bibitem{SitePref2020} J. K. Nørskov, T. Bligaard, and F. Abild-Pedersen, Rev. Mod. Phys. {\bf 92}, 045001 (2020).
\bibitem{Interface2025} Q. X. Zhang, Y. Zhu, X. F. Peng, W. G. Yang, Y. P. Le, and X. Xin, Phys. Scr. {\bf 100}, 065001 (2025).
\bibitem{StrainFinite2025} D. Moldarev, T. Tsakiris, R. Holeňák, E. Ntemou, M. Wolff, and D. Primetzhofer, APL Mater. {\bf 13}, 061118 (2025).
\bibitem{UniaxialStrain2015} G. Andersson, K. Aits, and B. Hjörvarsson, Sci. Rep. {\bf 5}, 10301 (2015).
\bibitem{EarlyThin2002} G. Andersson, K. Aits, and B. Hjörvarsson, J. Alloys Compd. {\bf 334}, 14 (2002).
\bibitem{Fukai2003} Y. Fukai and H. Sugimoto, J. Phys.: Condens. Matter {\bf 15}, 1389 (2003).
\bibitem{Keilbar2009} C. Keilbar and W. Wolf, Phys. Rev. B {\bf 80}, 075108 (2009).
\bibitem{Wang2011} F. Wang et al., J. Phys.: Condens. Matter {\bf 23}, 135402 (2011).
\bibitem{Sugimoto2013} H. Sugimoto and Y. Fukai, J. Phys.: Condens. Matter {\bf 25}, 115401 (2013).
\bibitem{Andersson2010} G. Andersson, K. Aits, and B. Hjörvarsson, J. Alloys Compd. {\bf 504}, S428 (2010).
\bibitem{ThinFilm2011} K. Aits et al., Phys. Rev. B {\bf 84}, 064101 (2011).
\bibitem{HydrogenSite2019} J. \AA. Gripsrud et al., J. Alloys Compd. {\bf 786}, 104 (2019).
\bibitem{ThinFilmReview2015} V. A. Baturin et al., Phys. Met. Metallogr. {\bf 116}, 843 (2015).
\end{thebibliography}
\end{document}